# SDL based Validation of a Node Monitoring Protocol

Anandi Giridharan and Pallapa Venkataram*

Indian Institute of Science, Department of ECE, INDIA
anandi@ece.iisc.ernet.in, pallapa@ece.iisc.ernet.in

## ABSTRACT

*Mobile ad hoc network is a wireless, self-configured, infrastructureless network of mobile nodes. The nodes are highly mobile, which makes the application running on them face network related problems like node failure, link failure, network level disconnection, scarcity of resources, buffer degradation, and intermittent disconnection etc. Node failure and Network fault are need to be monitored continuously by supervising the network status. Node monitoring protocol is crucial, so it is required to test the protocol exhaustively to verify and validate the functionality and accuracy of the designed protocol. This paper presents a validation model for Node Monitoring Protocol using Specification and Description Llanguage (SDL) using both Static Agent (SA) and Mobile Agent (MA). We have verified properties of the Node Monitoring Protocol (NMP) based on the global states with no exits, deadlock states or proper termination states using reachability graph. Message Sequence Chart (MSC) gives an intuitive understanding of the described system behavior with varying node density and complex behavior etc.*

## 1. INTRODUCTION

Node Monitoring is one the important task of fault management in networks, where Mobile Agents have proved that they are very efficient in node monitoring[1]. The usage of Mobile Agents gives the solution to the scalable problem in centralized network management[2]. Mobile Agents plays a vital role in node monitoring process. Agents carry out management function in an autonomous and efficient way[3]. This paper presents a formal model of the Node Monitoring Protocol based on SDL using the Finite State Model. Formal description using SDL specifies the functional operation of the protocol and also helps in detecting design errors like deadlock, livelock, unspecified reception, non-executable interactions, etc. The rest of the paper is organized as follows. Section 2 discusses on Significance of Node Monitoring Protocol (NMP) in Ubiquitous environment. Section 3 presents Formal SDL specification of NMP. Section 4. illustrates validation of NMP for various design errors like deadlock, unspecified reception, livelocks, etc. Section 5 presents Validation results of NMP using reachability graph. Section 6 draws the conclusion.

## 2. SIGNIFICANCE OF NODE MONITORING PROTOCOL IN UBIQUITOUS ENVIRONMENT

In a Ubiquitous Network, accurate and efficient monitoring of dynamically changing environment is very important in order to obtain the seamless transparency within mobile devices[4]. Monitoring resource allocation scheme for the Unodes, i.e., nodes running a ubiquitous application in a ubiquitous network is very important to check their Quality of Service. Static and Mobile Agent, based technology can provide a good framework to develop monitoring systems for ubiquitous network environment, since it can do complicated works on behalf of a node independently and transparently[5]. Static Agent sends a request to Mobile

Agent to collect raw resource information from the nodes like some of the health conditions like node failure, link failure, misbehaviour of the nodes in the network and to report the monitored results to them. Solution for entering the recovery upon validation is worked out that maintains the health of Node Monitoring Protocol[6].

## 2.1. Finite State Machine Formalism of Node Monitoring Protocol

An Finite State Machine M, is a 5-tuple A=(I, O, S, T, F) I is the Input, O is the output and S is the states and F is the finite sets. The main system which runs at the central node, where Static Agent is deployed for collection of network status information. The Mobile monitoring system is status monitoring segment, which runs in the migrated Mobile Agents. Figure 1. shows the State transition sequence that illustrates that NMP is capable of delivering data without duplication and in right order. Initially Static Agent which resides in the main segment in idle state then if requests arise, creates Mobile Agent and dispatches sending request $M_{req}$ to monitor the status of the node, initiating the timer. Even if channel loses $M_{req}$, time out occurs triggering retransmission. and time channel correctly delivers the message. Now Mobile agent sends Request to Node 1 and in case channel loses the Request, Time out occurs and retransmission of the data takes place. Request goes to Node 1 and Mobile Agent monitors the node collects the status of the Node like node failure, link failure, energy level, throughput etc, and delivers to the Static Agent and goes into idle state again. Many important properties of requirement specifications can be checked during requirements capture. First of all, requirements characterizing the total behavior of a system may be expressed in terms of temporal modalities (dynamic requirements) including safety and liveness conditions.

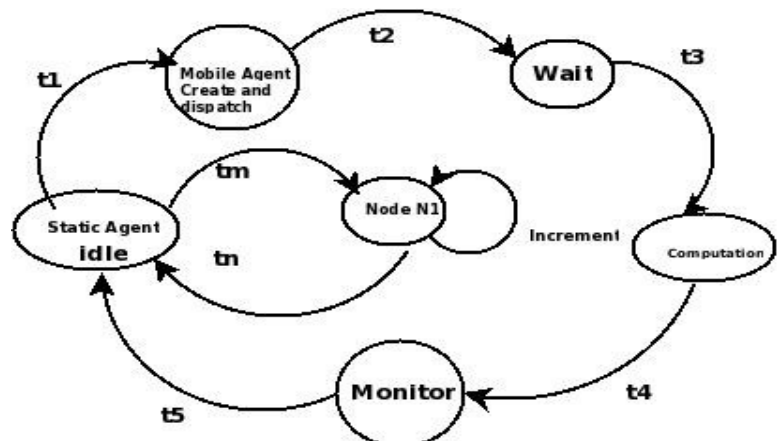

Figure 1: Formal FSM specification of NMP

## 2.2. NMP Functioning

**Liveness property:** In system verification and model checking, liveness properties are requirements that something good must eventually happen For example, with every request from Static Agent, Node status should be collected by Mobile Agent and protocol should terminate successfully.

**Safety property:** Bad things will not happen. For example. Node Monitoring Protocol should operate properly. MSC shows the behavior of the normal Node Monitoring Protocol as shown in the figure 2. We chose to rely on the FSM formalism because it suits very well to the analysis of data flows and allows to put constraints on the variables of the transitions.

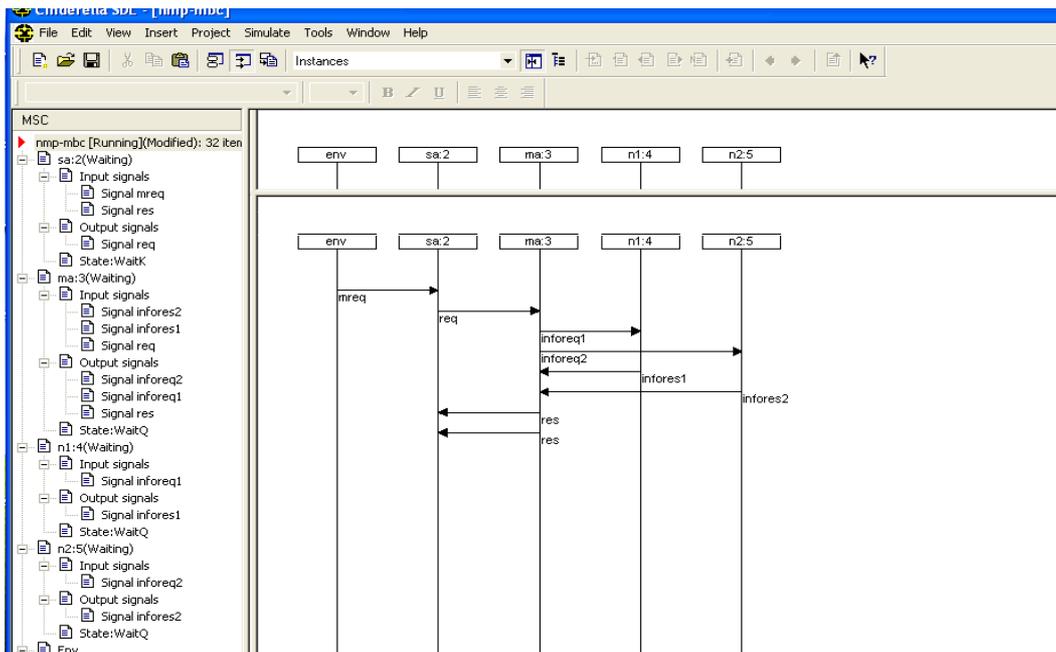

Figure 2: Message Sequence chart showing expected communication between various entities

## 2.3. Verification of NMP

We manually derived the EFSM directly from the IETF specification . The verification process consists to map the traces of I/O events (messages received and sent) recorded on each node, with the specification. As seen in Figure 3, $C_1$ is the outgoing channel of the Static Agent and $C_2$ is the outgoing channel of the Mobile Agent.

**Proof of Liveness Property**

Liveness property is taken care in design process, they include termination of the protocol. From above transition state, we observe that message $M_{req}$ and Response are transmitted from and to Static Agent respectively even under the conditions of frame and acknowledgement loss and NMP returns to its terminator state. Hence Specified messages have been transmitted and received correctly.

**Proof of Safely properties**

From transitions, we can see handling of lost frames and Acknowledgement are done by retransmission and no redundancy has occurred by sending two duplicates of the same message. Hence safety property.

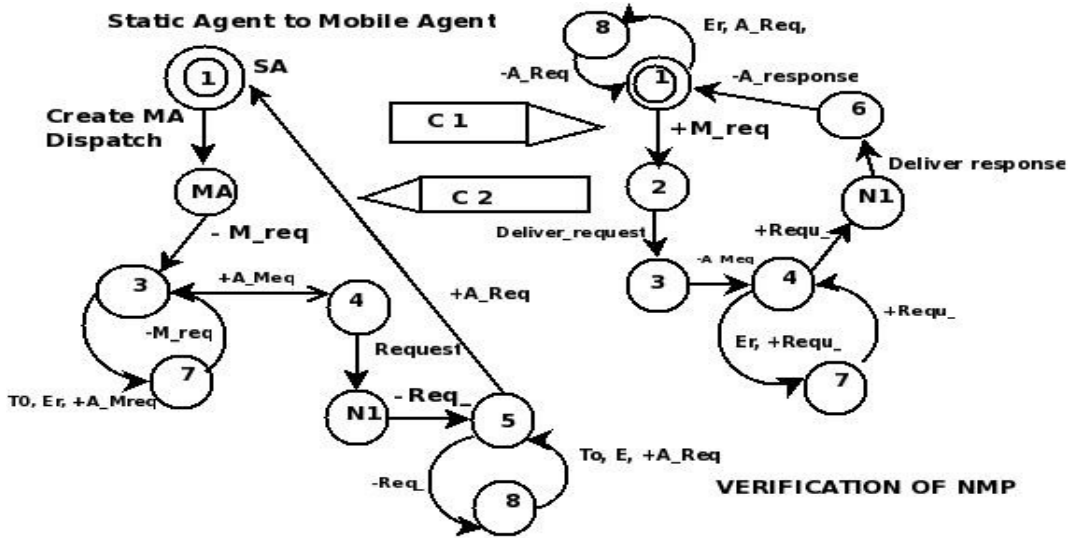
Figure 3: Verification of NMP

## 3. Formal Specification of Node Monitoring Protocol using SDL

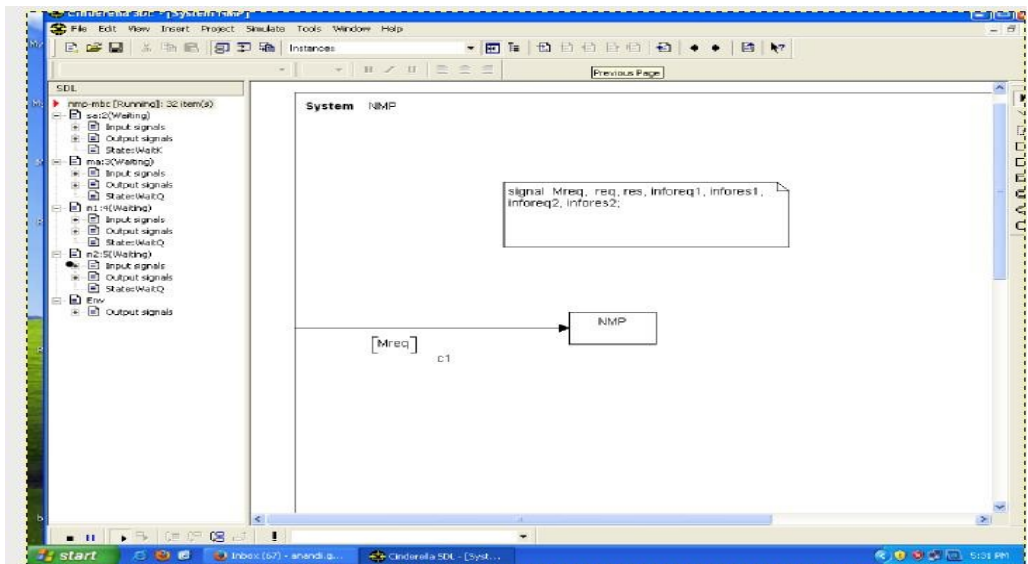
Figure 4: System of Node Monitoring protocol using SDL

We choose SDL (Specification and Description Language) as the target language because it supports more highly-detailed design, so that code automatically generated from the specification can be a much more complete implementation of the system compared to the code generated by UML. The syntax and semantics of SDL are formally defined, standardized, and maintained by the International Telecommunication Union. Its goal is to specify the behavior of a system from the representation of its functional aspects. The top level of an SDL specification is a system agent consisting of two sub-agents, Static agent and Mobile agent. Blocks of the node monitoring protocol are used to define a system structure as shown in figure 4. We have considered 4 blocks, Static Agent, Mobile Agent, Node 1 and Node 2 as shown in figure 5. Process specifies the behavior of a system from the representation of its functional aspects. We have shown the behavior of the processes of Mobile Agent process, Static Agent process, Node1 process and Node2 process as shown in figures 6, 7, 8 and 9 respectively. Signal routes transfer signal immediately while channels may be delaying. The signal specification identifies the name of the signal type and the sorts of the parameters to be carried by the signal such as Mreq, resp, req, inforequest1, inforesponse1, inforequest2, inforesponse2. As seen in the SDL model, SA behavior is expressed as a process which exists in a state, waiting for an input (event) triggered from environment. When $M_{req}$ signal is sent from environment, SA locates Mobile Agent and sends a request to collect health of the nodes. We have considered two Nodes N1 and N2 in our case. Mobile Agent interacts with the nodes and gets node information back to Static Agent. We have to note that such a specification may contain few errors during its design even from the requirements. For this reason, we have used model checking technique like generating Message sequence chart in order to verify our specification. Indeed, before validating an implementation we need to make sure that the used specification corresponds to the requirements. Simulation was done to verify that specification is free from deadlocks and live-locks within simulated space. Presence of such dead-locks or livelocks reveals that Node monitoring protocol system does not behave as expected that can be monitored using Message sequence chart that is generated after simulation. MSCs are another valuable description technique for visualizing and specifying inter-system, asynchronous component interaction[7]. MSC strength lies in their ability to describe communication between cooperating processes. There are arrows representing messages passed from a sending to a receiving process.

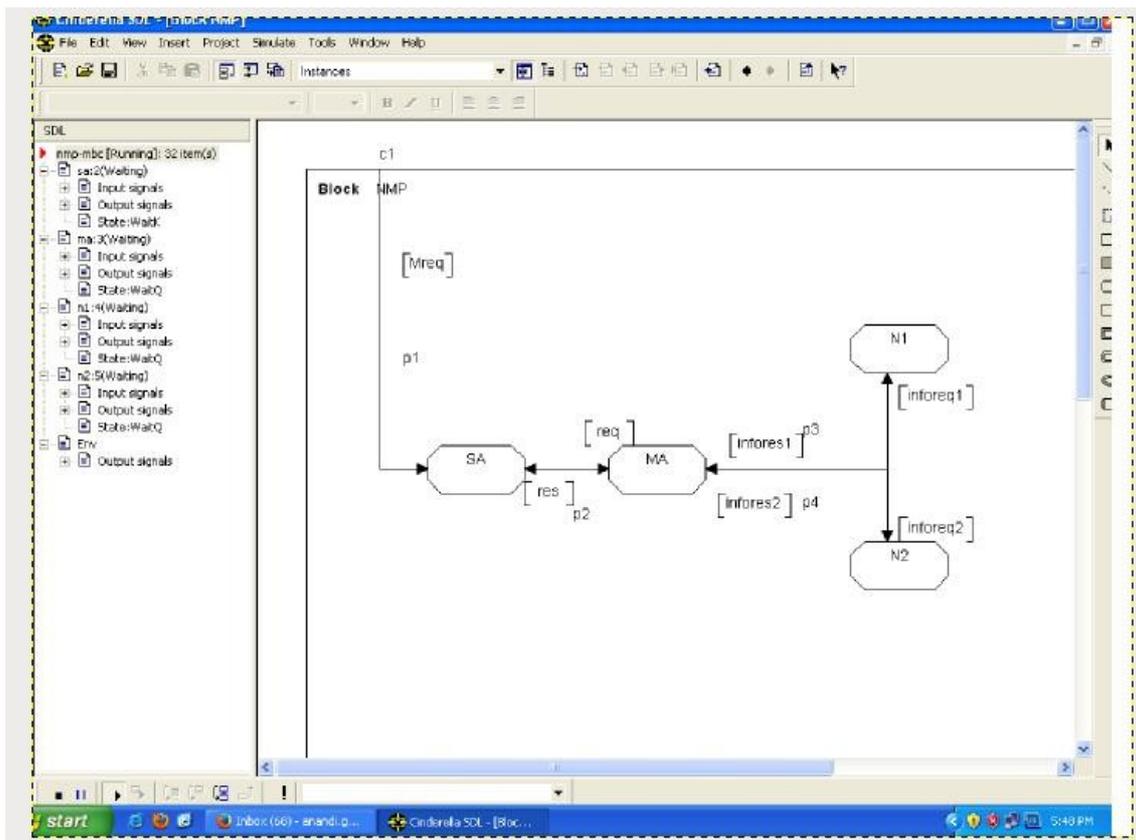

Figure 5: Blocks of Node Monitoring protocol using SDL

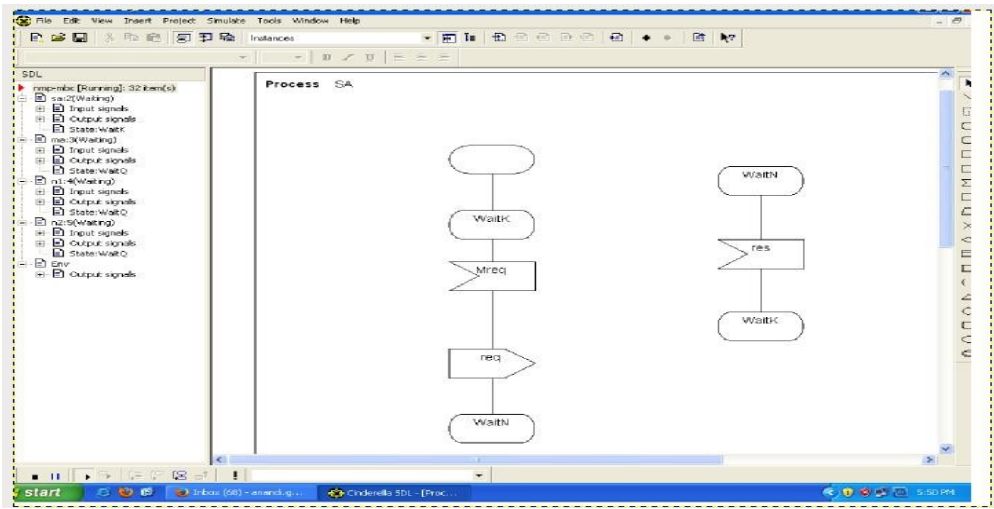

FIGURE 6: PROCESS STATIC AGENT OF NODE MONITORING PROTOCOL USING SDL

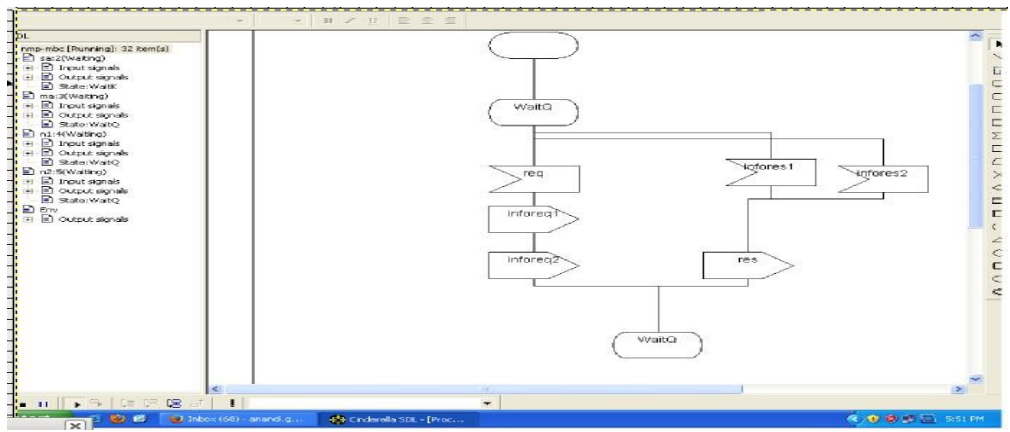

FIGURE 7: PROCESS MOBILE AGENT OF NODE MONITORING PROTOCOL USING SDL

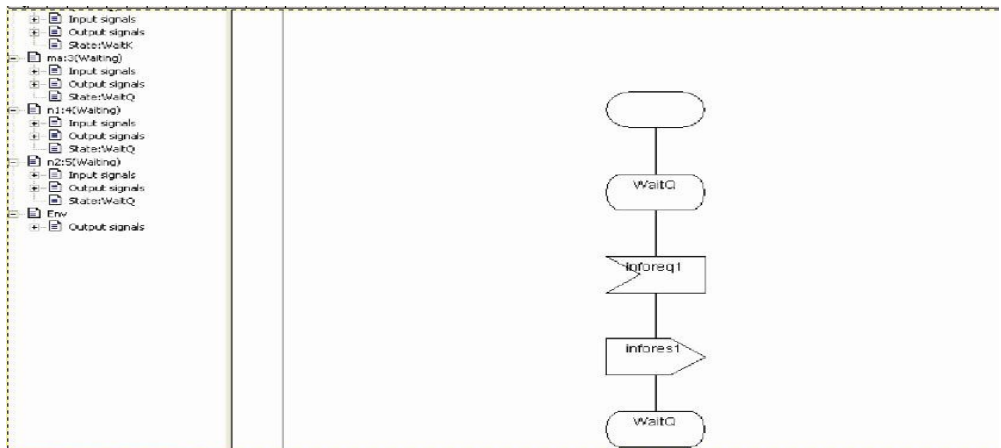

Figure 8: Process of Node 1 of Node Monitoring protocol using SDL

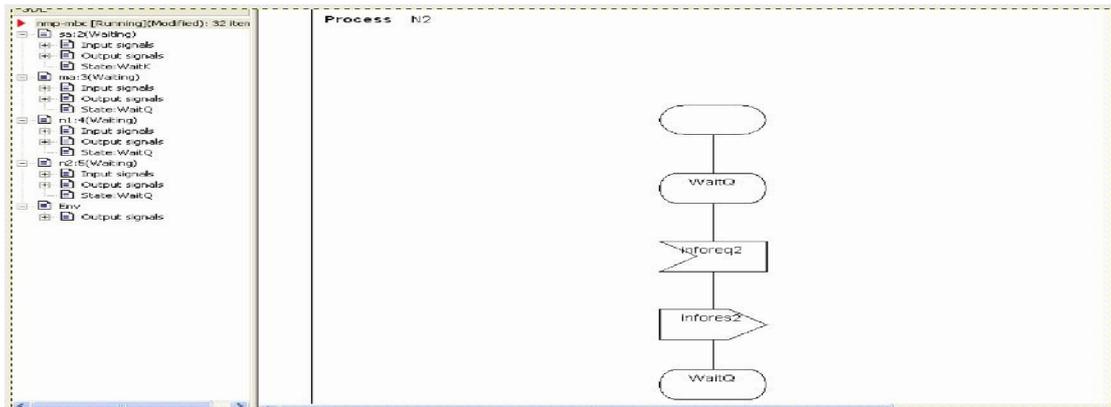
Figure 9: Process of Node 2 of Node Monitoring protocol using SDL

## 4. Validation of Node Monitoring Protocol

Failures may also arise at run-time, for example, because of the loss of network connectivity. node failure, link failure etc. The design of the framework must ensure its ability to hold good under increasing load, increasing complexity of requests and increasing size of resulting composite services[8]. Validation ensures that the protocol specifications will not get into protocol design errors. (Deadlock, unspecified reception, livelock etc). We have used Message sequence charts for validation of Node Monitoring Protocol. MSCs were used to identify different kinds of errors like Deadlock, unreachable states, livelocks etc.

### 4.1. Deadlock

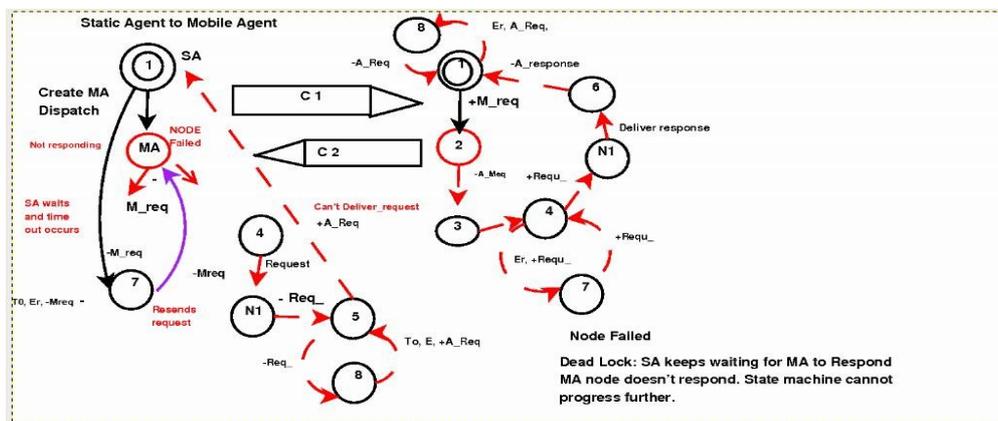
Figure 10: Deadlock error in Node Monitoring protocol

Deadlock is a scenario, whereby state machines cannot progress to another state because they are waiting for an event that will never occur. Static Agent sends creates Mobile Agent and dispatches, due to the failure of the node, Mobile Agent does not respond to the request of Static Agent. Static Agent waits for random time and time out occurs and again sends request to Static Agent and again goes to wait state. So both the state machines cannot progress further waiting for event to occur that never happens. Hence Deadlock occurs as seen in figure 10**.** Referring to the Message sequence chart, we can see that Static agent send the request to Mobile Agent. Due to failure of Mobile Agent node, it does not respond. Static agent waits for certain time. Time out occurs and again new request is sent from Static Agent and again goes to wait process expecting Mobile agent to respond, which does not happen. So state machines cannot progress further waiting for event to occur, that never happens. Hence Dead lock occurs. Figure 11 shows the MSC of NMP that indicates Deadlock, where 2 process cannot progress further waiting event to occur.

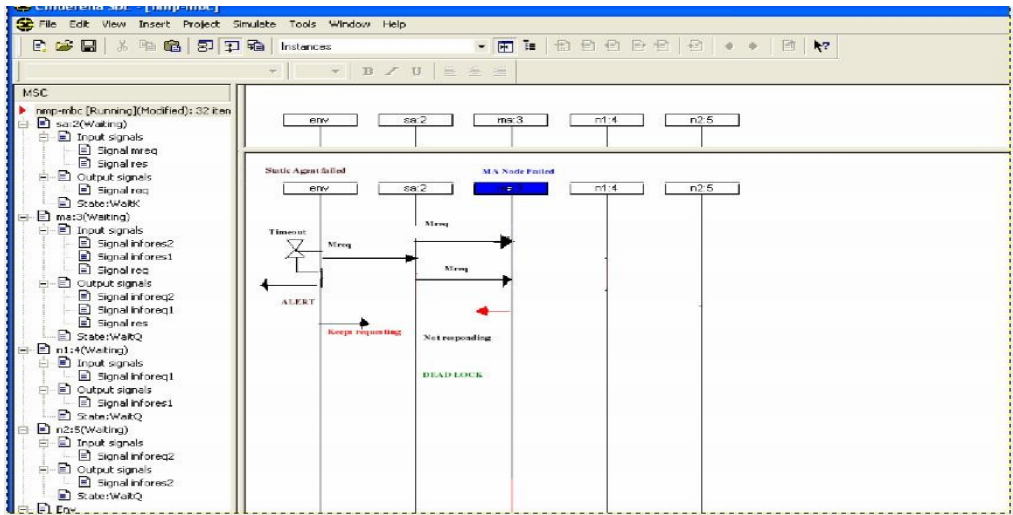
Figure 11: Message Sequence chart showing Dead Lock error in NMP

### 4.2 Unspecified Reception:

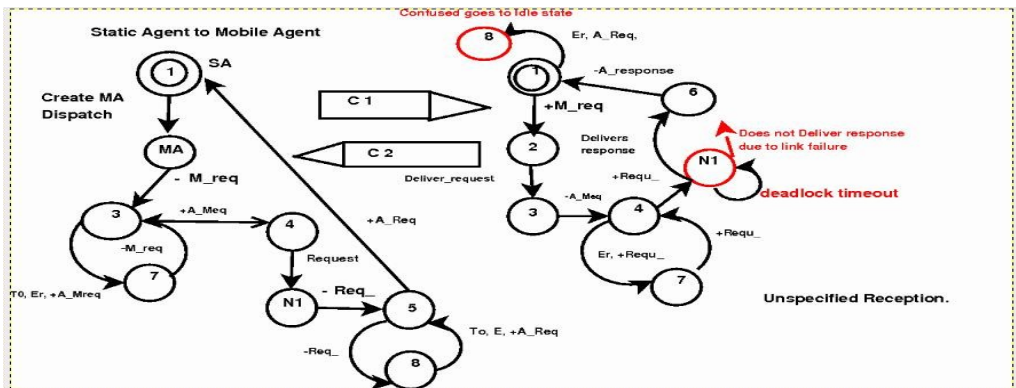
Figure 12: Unexpected State error in NMP

Use of timers may prevent deadlocks, but their use may result in states that are never reached if the specification is faulty[9]. In our simulation, When there was no request from environment, Static Agent is in idle state. Once the request comes from environment , Static Agent sends request to Mobile Agent. Mobile agent goes to Nodes and collects their status. In this case error will propagate because a generic deadlock timer expired that was unaware of the state specific actions to take at this point. So due to ambiguity, Static Agent is not in position to decide what state it should be, hence goes idle. Even through Mobile Agent is ready with node status, Static Agent is not a possible to accept the information as shown in figure 12. Figure 13. shows the MSC indicating the unexpected state error due to ambiguity. Figure 12: Unexpected State error in NMP

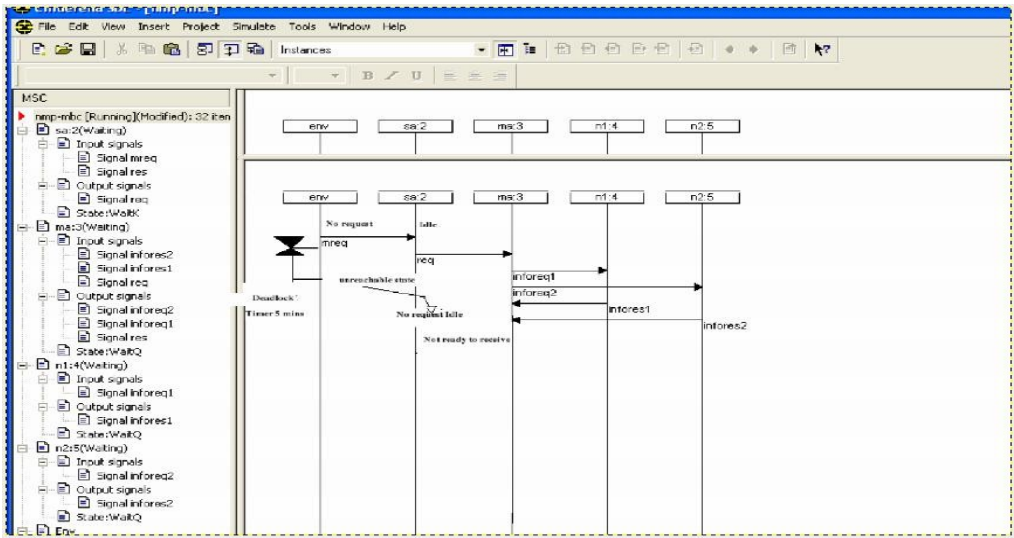
Figure 13: Message Sequence chart showing Unexpected State error in NMP

### 4.3. Data loss:

As indicated in figure 14, request from Node1 gets lost in channel and no response from Mobile Agent regarding status of the Node. Figure 15. shows that Request sent by Static Agent to Mobile Agent and request gets lost in the channel, Response comes from only from Node2 to Mobile Agent. Data loss occurs, when one or more packets of data travelling across a network fail to reach their destination. Data loss can be caused by a number of factors, including packet drop because of channel congestion, rejected corrupted packets, faulty networking hardware. As seen in the figure 16, it shows that the data loss increases if more number of packets are sent. Hence throughput will be less due to the number of retransmission.

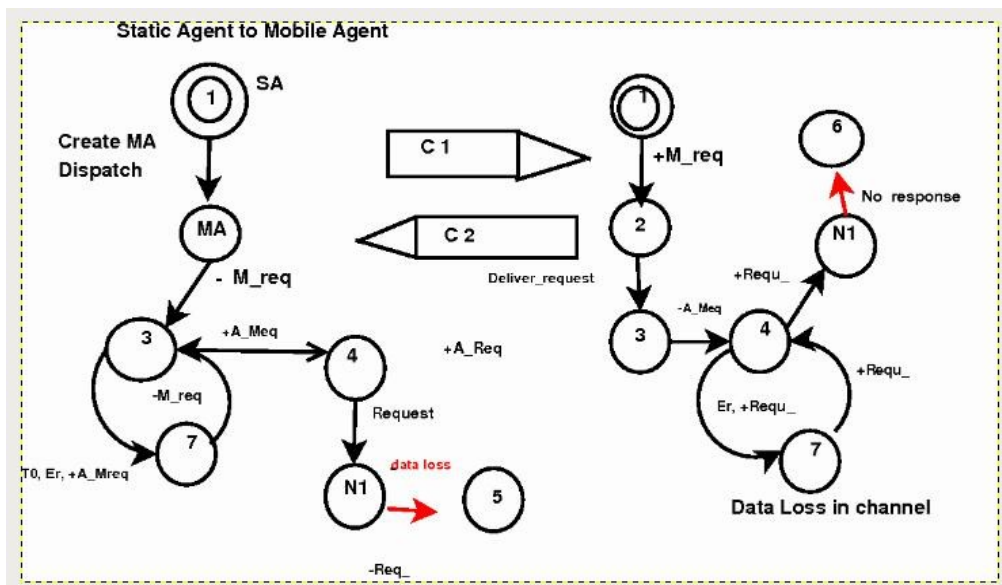
Figure 14: Data Loss occurring in channel

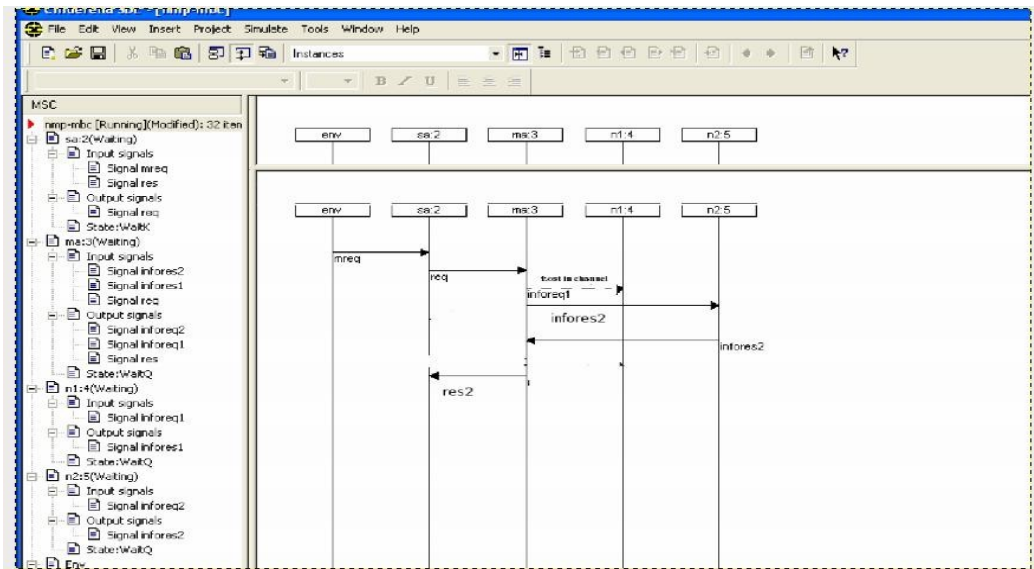
Figure 15: Message Sequence chart showing Data Loss occurring in channel

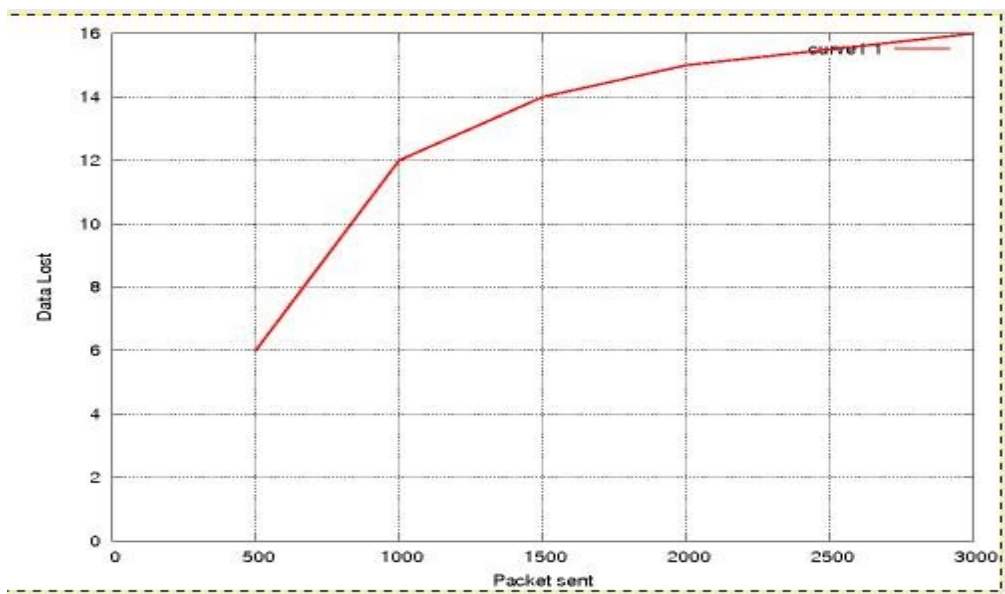
Figure 16: Data loss verses number of packets sent

### 4.4 Livelocks:

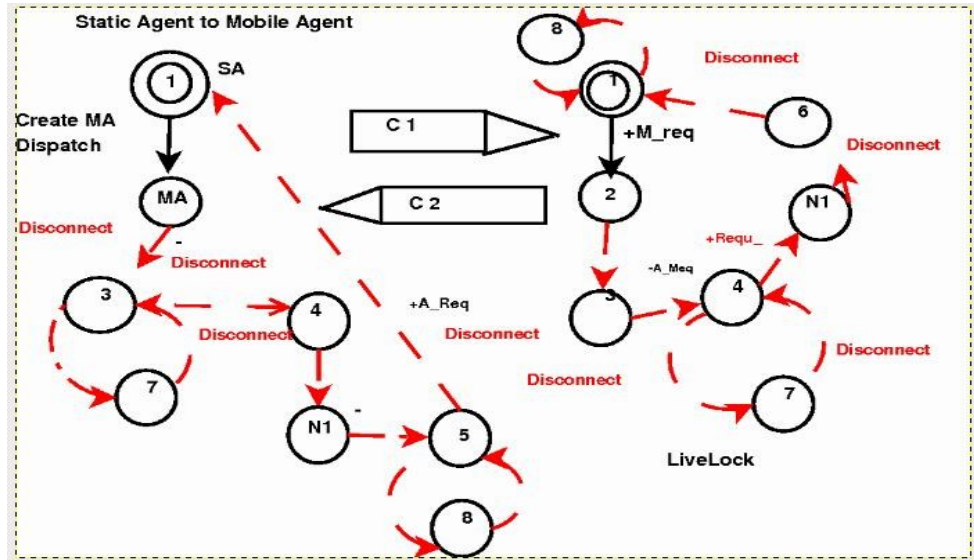

Figure 17: Message Sequence chart shows infinite loop livelock error

Livelock is a scenario whereby sequences of messages is repeated in an endless loop as shown in figure 17. Without appropriate safety mechanisms livelock can consume all of the resources in a network. Livelocks can occur depending on the value of data, such as an entity forwarding a message to itself . MSC indicates, how sequence of messages are repeated in an endless manner as shown in figure 18.

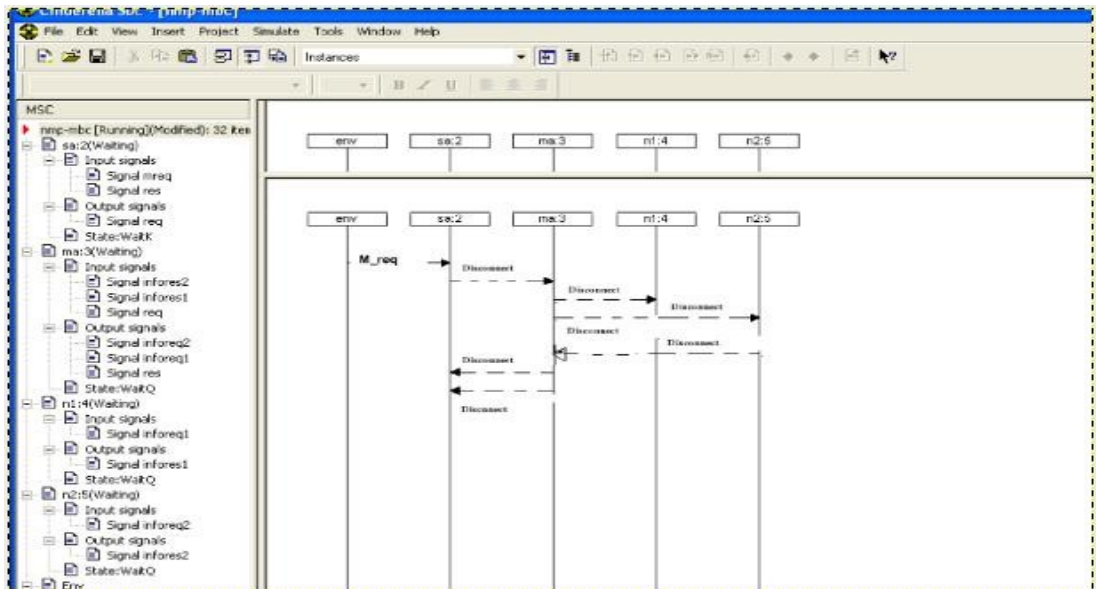

Figure 18: Message Sequence chart shows infinite loop livelock error

## 5. Validation Results of NMP using Reachability Graph

The most straightforward technique to validate a given network of two communicating FSMs is called state exploration. We have considered Node monitoring Network [Mobile Agent, Node] whose communicating FSMS sender machine and reaching machine are as shown in figure 19.. The exchanged messages between two machines have the following meaning:

-M $_{req}$ denotes a request sent to Mobile Agent from environment.
+M req denotes Postive acceptance of M req.
-Req denotes a request sent from Mobile Agent to Node.
+Req denotes postive acceptance of request from Node.
-A req Acknowledgement sent from Mobile Agent.
+A req Positive Acknowledgement from Node to Mobile Agent.

In order to describe the behavior of our network, many processes have been specified and tested. Specification may contain errors like deadlock, unspecified reception, data loss etc. For this reason, using reachablility analysis, we had to validate our specification. During validating FSM, we verified that specification had errors like deadlock, unspecified reception and one process terminated successfully.

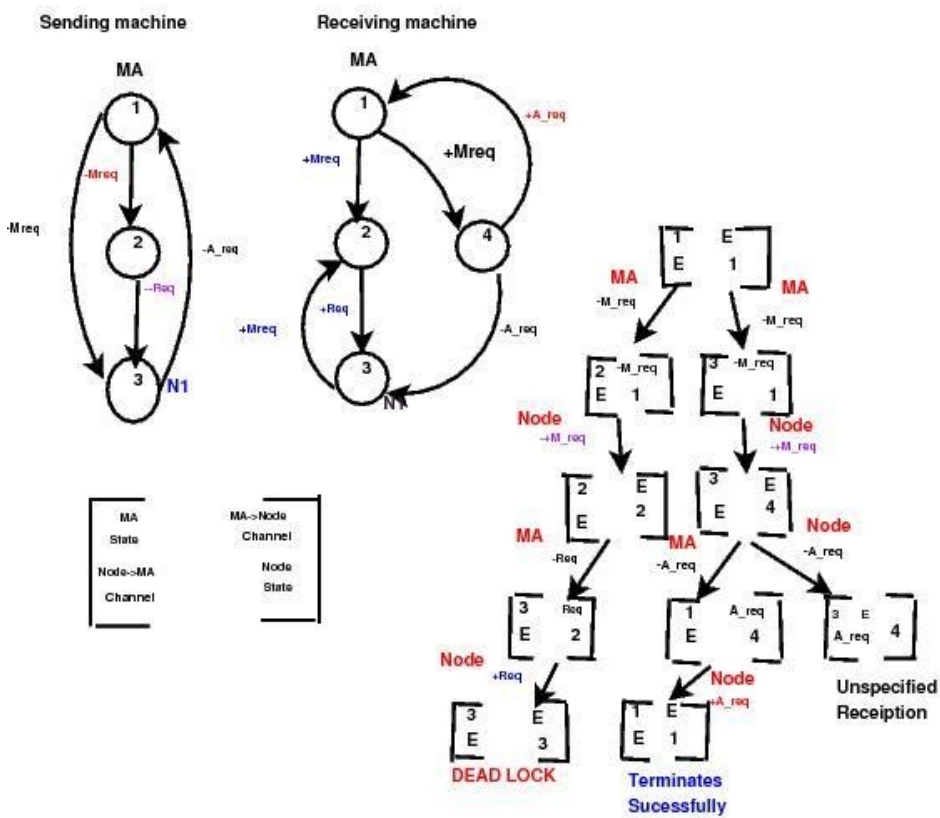

Figure 19: Reachability Graph for Node Monitoring Protocol

## 6. Simulation and Results

Simulation was used in both protocol specification and validation using Cinderella SDL tool to conduct verification and validation of Node Monitoring Protocol. We simulated on five to fifteen nodes. It was found that various errors increases as the traffic on the network increased. Simulation results on data loss, deadlock error, unspecified error and performance of the protocol are summarized below..

### 6.1 Dead Lock

Deadlocks occur when two or more processes interfere with each other in such a way that the network as a whole eventually cannot proceed. Multiple processes, and multiple processes have always given rise to deadlocks of various kinds. Graph 20. shows Dead-lock error rate versus Number of processes. As the number of process increased, the deadlock error also increases.

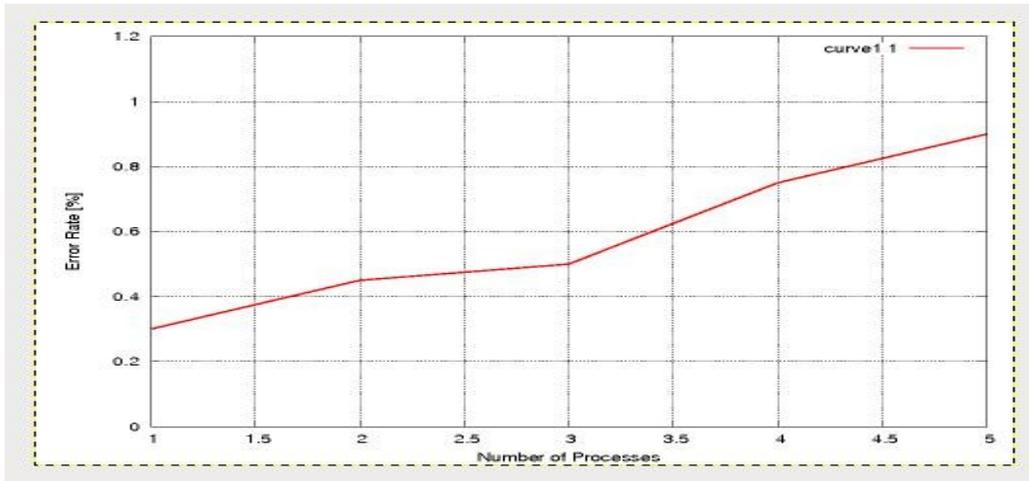

Figure 20: Deadlock error verses number of process

### 6.2 Unspecified Error

Simulation was conduction on cinderella SDL tool for 5, 10 and 15 nodes, we found that as the unspecified error increases delay increases as seen in the figure 21 and also we found that as the number of nodes increased unspecified reception error also increased as seen in figure 22.

### 6.3 NMP performance

NMP performance is an overall measure of the effciency of the system's achievement in terms of rates and throughputs. The results of applying variation in data transmissions versus Error rate are drawn in figure 23. with two sets of settings. Curve 1 and 2 were for heavy data transmission rate and slower data transmission rate. The larger data transmission rate, more the error rate. This is due to messages are lost in the network. On the other hand, decreasing date rate below certain value will discard reply messages that may arrive a bit later. The best choice of data transmission is that one with less errors as shown in figure 24 . It is obvious that performance improves only when the bottleneck transition time is decreased.

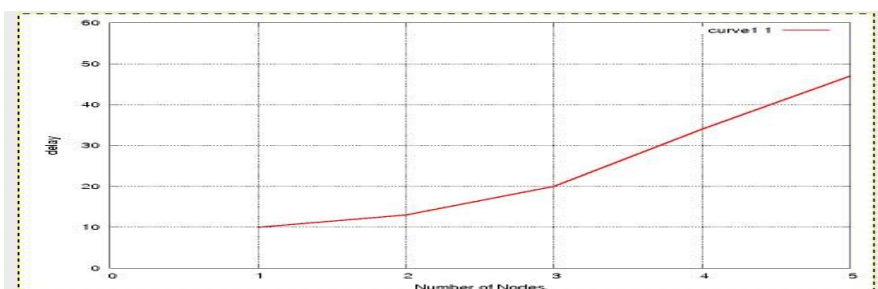

.Figure 21: Delay verses number of Nodes

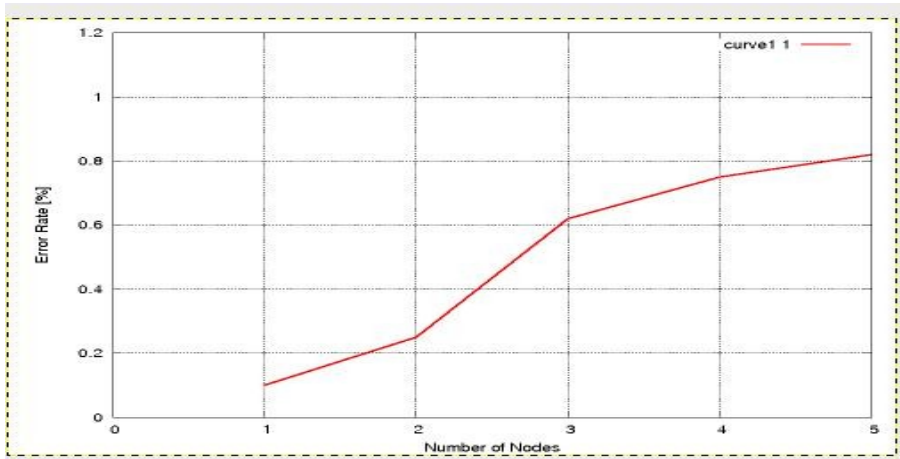

Figure 22: Deadlock error vs number of Nodes

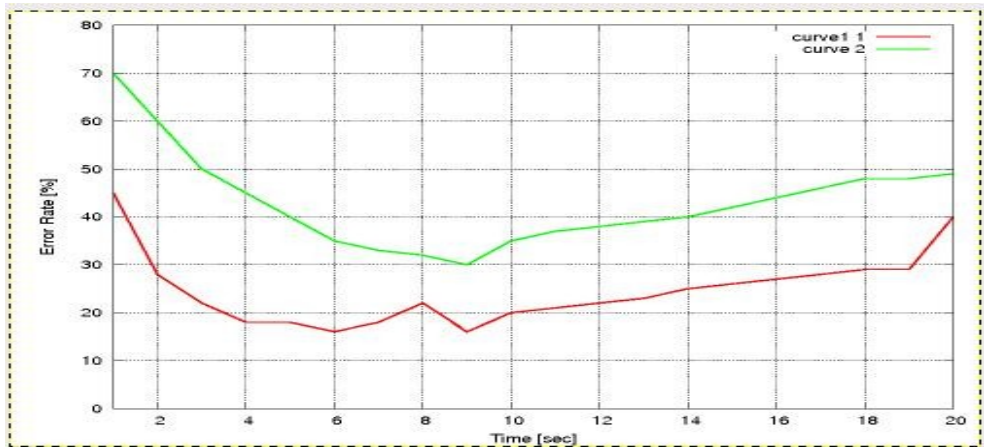

Figure 23: High Data transmission rates versus Error Rate

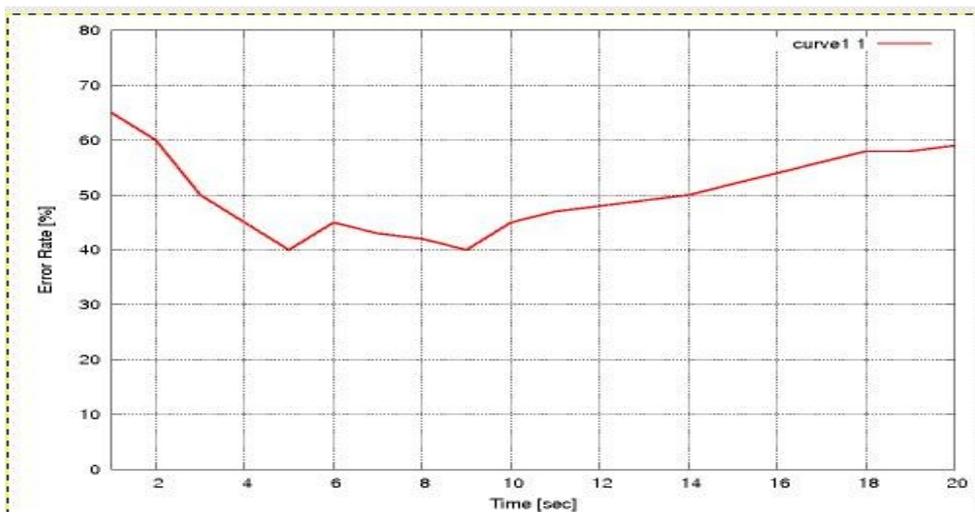

Figure 24: Light Data transmission rate versus Error Rate

## 7. CONCLUSION

This paper has presented verification and validation model for the Node Monitoring protocol. It includes a formal specification of the protocol using Specification and Description Language and Message sequence charts a method and a tool for the automated test generation of scenarios. Validation checks for safety and liveness property of the protocol to check proper functioning and termination of protocol and validation model presents several advantages[10][11]. Reachability analysis was carried out to check the correctness properties of NMP. First, the design of a formal specification from which tests are generated contributes to eliminate design errors like Deadlock, unspecified receptions and livelocks and using SDL, it is shown that design flaws and ambiguity introduced in informally specified, textual protocols can be avoided if protocol is formally modelled.

**Authors**

**First Author.**

Anandi Giridharan, received MSc(Engg) from Indian Institute of Science. She currently working as Senior Scientific Officer in ECE Department, Indian Institute Science, Bangalore. Her Research Interest are in area of Ubiquitous Learning, Communication Protocols and Multimedia systems.

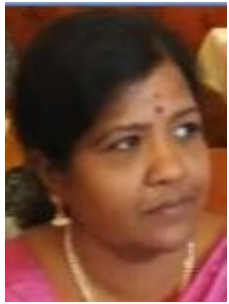

**Second Author**

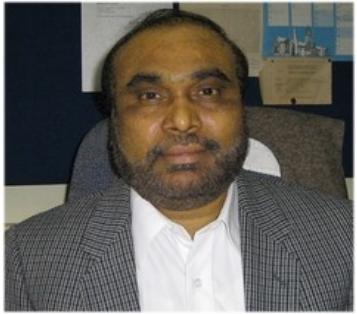

Prof. Venkataram Pallapa received his Ph.D. Degree in Information Sciences from the University of Sheffield, England, in 1986. He is currently the chairman for center for continuing education, and also a Professor in the Department of Electrical Communication Engineering, Indian Institute of Science, Bangalore, India.

Dr. Pallapa's research interests are in the areas of Wireless Ubiquitous Networks, Communication Protocols, Computation Intelligence applications in Communication Networks and Multimedia Systems.

Dr. Pallapa is the holder of a Distinguished Visitor Diploma from the Orrego University, Trujillo, PERU. He has published over 150 papers in International/national Journals/conferences. Written three books: Mobile and wireless application security, Tata McGraw-Hill, Communication Protocol Engineering, publications Prentice-Hall of India (PHI), New Delhi, 2014 (Co-author: Sunil Manvi, B Satish Babu) and Multimedia: Concepts & Communication, Darling Kinderley(India) Pvt. Ltd., licensees of Pearson Education in South Asia, 2006. Written chapters for two different books, and a guest editor to the IISc Journal for a special issue on Multimedia Wireless Networks. He has received best paper awards at GLOBECOM'93 and INM'95 and also CDIL (Communication Devices India Ltd) for a paper published in IETE Journal. He is a Fellow of IEE (England), Fellow of IETE(India) and a Senior member of IEEE Computer Society.